\documentstyle[aas2pp4,psfig]{article}
\def\tauc{\ifmmode \tau_{\rm c}\else$\tau_{\rm c}$\fi}
\def\tpsr{\ifmmode t_{\rm PSR}\else$t_{\rm PSR}$\fi}
\def\twd{\ifmmode t_{\rm WD}\else$t_{\rm WD}$\fi}
\def\Mcomp{\ifmmode M_{\rm C}\else$M_{\rm C}$\fi}
\def\Rcomp{\ifmmode R_{\rm C}\else$R_{\rm C}$\fi}
\def\Lcomp{\ifmmode L_{\rm C}\else$L_{\rm C}$\fi}
\def\Mpsr{\ifmmode M_{\rm PSR}\else$M_{\rm PSR}$\fi}
\def\Msun{\ifmmode M_\odot\else$M_\odot$\fi}
\def\Rsun{\ifmmode R_\odot\else$R_\odot$\fi}
\def\Teff{\ifmmode T_{\rm eff}\else$T_{\rm eff}$\fi}
\def\kms{\ifmmode {\rm km\,s^{-1}}\else$\rm km\,s^{-1}$\fi}
\def\mv{\ifmmode m_{\rm F555W}\else$m_{\rm F555W}$\fi}
\def\mi{\ifmmode m_{\rm F814W}\else$m_{\rm F814W}$\fi}
\def\Mv{\ifmmode M_{\rm F555W}\else$M_{\rm F555W}$\fi}
\def\Mi{\ifmmode M_{\rm F814W}\else$M_{\rm F814W}$\fi}
\def\psr{PSR~B1855+09}
\def\Sref#1{\S\,\ref{sec:#1}}
\let\simgt\gtrsim
\let\simlt\lesssim
\let\internalcite\cite
\def\cite{\def\citename##1{##1}\internalcite}
\def\ctyr{\def\citename##1{}\internalcite}
\makeatletter
{\pt@width\wd\pt@box\box\pt@box\spew@ptblnotes%
\typeout{Page \the\pt@page\space of table \thetable\space has been set to
width \the\pt@width\space with \the\pt@nlines\space lines per page}%
\endcenter\end@float}
\def\startdata{\pt@line=0\pt@calcnlines%
\ifdim\pt@width>\z@\def\@halignto{to \pt@width}\else\def\@halignto{}\fi%
\let\fnum@table=\fnum@ptable\set@mkcaption%
\@float{table}\center\caption{\pt@caption}\leavevmode%
\setbox\pt@box=\pt@tabular{\pt@format}\pt@head}
\newenvironment{deluxetable*}[1]{\def\pt@format{\string#1}%
\set@tblnotetext\global\pt@ncol=0\global\pt@column=0\global\pt@page=1%
\def\pt@addcol{\global\advance\pt@ncol by\@ne}}%
{\pt@width\wd\pt@box\box\pt@box\spew@ptblnotes%
\typeout{Page \the\pt@page\space of table \thetable\space has been set to
width \the\pt@width\space with \the\pt@nlines\space lines per page}%
\endcenter\end@dblfloat}
\def\startdata{\pt@line=0\pt@calcnlines%
\ifdim\pt@width>\z@\def\@halignto{to \pt@width}\else\def\@halignto{}\fi%
\let\fnum@table=\fnum@ptable\set@mkcaption%
\@dblfloat{table}\center\caption{\pt@caption}\leavevmode%
\setbox\pt@box=\pt@tabular{\pt@format}\pt@head}
\def\thebibliography{\subsection*{REFERENCES}
\list{}{\labelwidth3em\leftmargin\labelwidth\labelsep\z@\parsep\z@
\itemsep\z@\itemindent-3em\usecounter{enumi}}
\def\refpar{\relax}
\def\newblock{\hskip .11em plus .33em minus .07em}
\sloppy\clubpenalty4000\widowpenalty4000
\sfcode`\.=1000\relax}
\makeatother

\begin{document}

\title{The Temperature and Cooling Age of the White-Dwarf Companion to the
       Millisecond Pulsar PSR~B1855+09}

\righthead{Temperature and Cooling Age of the Companion to \psr}

\author{M. H. van Kerkwijk\altaffilmark{1},
        J. F. Bell\altaffilmark{2},
        V. M. Kaspi\altaffilmark{3}, 
	S. R. Kulkarni\altaffilmark{4}
\altaffiltext{1}{Astronomical Institute, Utrecht University, P. O. Box
                 80000, 3508 TA~~Utrecht, The Netherlands;
                 M.H.vanKerkwijk@astro.uu.nl}
\altaffiltext{2}{Australia Telescope National Facility, CSIRO, PO Box
                 76, Epping 1710, NSW, Australia; jbell@atnf.csiro.au}
\altaffiltext{3}{Massachusetts Institute of Technology, Physics
		 Department, Center for Space Research 37-621, 70
		 Vassar Street, Cambridge, MA 02139;
		 vicky@space.mit.edu}
\altaffiltext{4}{Palomar Observatory, California Institute of
                 Technology 105-24, Pasadena, CA 91125, USA; 
                 srk@astro.caltech.edu}
}

\begin{abstract} 
We report on Keck and {\em Hubble Space Telescope} observations of the
binary millisecond pulsar PSR~B1855+09.  We detect its white-dwarf
companion and measure $\mv=25.90\pm0.12$ and $\mi=24.19\pm0.11$ (Vega
system).  From the reddening-corrected color,
$(\mv-\mi)_0=1.06\pm0.21$, we infer a temperature
$\Teff=4800\pm800\,$K.  The white-dwarf mass is known accurately from
measurements of the Shapiro delay of the pulsar signal,
$\Mcomp=0.258^{+0.028}_{-0.016}\,\Msun$.  Hence, given a cooling
model, one can use the measured temperature to determine the cooling
age.  The main uncertainty in the cooling models for such low-mass
white dwarfs is the amount of residual nuclear burning, which is set
by the thickness of the hydrogen layer surrounding the helium core.
From the properties of similar systems, it has been inferred that
helium white dwarfs form with thick hydrogen layers, with mass
$\simgt\!3\times10^{-3}\,\Msun$, which leads to significant additional
heating.  This is consistent with expectations from simple
evolutionary models of the preceding binary evolution.  For
PSR~B1855+09, though, such models lead to a cooling age of
$\sim\!10\,$Gyr, which is twice the spin-down age of the pulsar.  It
could be that the spin-down age were incorrect, which would call the
standard vacuum dipole braking model into question.  For two other
pulsar companions, however, ages well over 10\,Gyr are inferred,
indicating that the problem may lie with the cooling models.  There is
no age discrepancy for models in which the white dwarfs are formed
with thinner hydrogen layers ($\simlt\!3\times10^{-4}\,\Msun$).
\end{abstract}


\section{Introduction}\label{sec:intro}

About one twentieth of the known radio pulsars reside in binary
systems.  For most of these, the companions have estimated masses
between 0.1 and 0.4\,\Msun\ and are thought to be low-mass, helium
white dwarfs.  Generally, the properties of the pulsars in these
systems differ markedly from those of typical isolated pulsars,
showing more rapid spin periods and smaller inferred magnetic fields
(for reviews, see, e.g., \cite{pk94}; \cite{vdh95}).  Presumably, this
is because of a phase of mass and angular momentum transfer which
occurred when the progenitor of the current white dwarf ascended on
the giant branch and overfilled its Roche lobe.  At the cessation of
mass transfer, the neutron star turned on as a millisecond radio
pulsar and the companion was left as a helium white dwarf.

An interesting property of these binaries is that they contain two
independent clocks that started running more or less simultaneously:
the spin-down age of the pulsar and the cooling age of the white dwarf
(\cite{k86}).  Assuming that the pulsar spins down due to a torque
$N\propto\nu^n$, the pulsar age is given by
\begin{equation}
\tpsr = \frac{P}{(n-1)\dot{P}} 
        \left[1-\left(\frac{P_0}{P}\right)^{n-1}\right],
\label{eq:tpsr}
\end{equation}
where $P\equiv1/\nu$ is the current spin period, $\dot{P}$ is its rate
of change, $P_0$ is the period when the pulsar began spinning down
following cessation of mass transfer, and
$n=\nu\ddot{\nu}/\dot{\nu}^2$ is the ``braking index,'' equal to 3
under the assumption of magnetic dipole radiation (for a review, see
\cite{ls98}).  For $n=3$ and $P_0\ll{}P$,
$\tpsr\simeq\tauc\equiv{}P/2\dot{P}$, where \tauc\ is the pulsar
``characteristic age.''

The cooling age of the white dwarf, the second clock, can be
determined from the stellar temperature and mass using a cooling
model.  The cooling properties of helium white dwarfs have been
modeled extensively, particularly since the optical identification of
a number of white-dwarf companions of millisecond pulsars.  A central
issue that has arisen is how much hydrogen remains when the white
dwarf is formed.  Most likely, the amount of hydrogen left will be
anti-correlated with core mass, since for larger core masses the
pressure at the core-envelope interface required for CNO cycle shell
burning can be maintained down to lower envelope masses.  If a white
dwarf is left with a sufficiently thick remaining hydrogen layer, with
mass $\simgt\!10^{-3}\,\Msun$, residual nuclear burning in the p-p
cycle will be a significant source of heat during the further
evolution, and, for a given age, the white dwarf will be hotter than
one whose hydrogen layer is thinner.

Evidence in favor of thick hydrogen layers and significant nuclear
burning comes from the millisecond pulsar binary system
PSR~J1012+5307.  For this binary, $\tpsr\simeq7\,$Gyr (\cite{lfln95}),
while from the surface temperature and gravity, as determined using
optical spectroscopy ($\Teff\simeq8500\,$K, $\log{}g\simeq6.7$;
\cite{vkbk96}; \cite{cgk98}), one infers a much shorter cooling age,
$\twd\simeq0.5\,$Gyr, if one uses models in which the white dwarf is
assumed to have formed with a relatively thin hydrogen layer
($\simlt\!3\times10^{-4}\,\Msun$), with, in consequence, little
nuclear burning (\cite{lfln95}; \cite{smc96}; \cite{ab97};
\cite{hp98}).  It has been argued that this implies that the pulsar
has not yet spun down much, i.e., $P_0\simeq{}P$.  Alberts, Savonije,
\& Van den Heuvel (\ctyr{ash96}) and Driebe et al.\ (\ctyr{dsbh98}),
however, find from simple evolutionary models in which the mass loss
during the red giant phase is simulated, that the helium white dwarf
companion should have had a much thicker hydrogen layer at formation,
$\sim\!5\times10^{-3}\,\Msun$.  With this thicker layer, they obtain a
cooling age that is consistent with the pulsar spin-down age.

The results are less clear for other systems with optically identified
companions, since these are too faint for optical spectroscopy and
their masses can only be estimated from the mass function (using a
guess for the pulsar mass and statistical arguments for the orbital
inclination).  There is one system, however, for which the mass of the
white dwarf is known precisely.  This system is \psr, composed of a
5.4\,ms radio pulsar and a low-mass companion in a 12.3\,d circular
orbit (\cite{srs+86}).  The pulsar is a very stable rotator and it has
been possible to measure the general relativistic Shapiro delay of the
pulsar signal near superior conjunction (\cite{rt91a}; \cite{ktr94}).
From the measurements, one infers that the binary is nearly edge on
($\sin{}i=0.9992^{+0.0004}_{-0.0007}$) and that the companion has mass
$\Mcomp=0.258^{+0.028}_{-0.016}\,\Msun$.  In addition, from the
measured parallax, $\pi=1.1\pm0.3\,$mas, the distance is
$0.91_{-0.20}^{+0.35}\,$kpc.  The characteristic age \tauc\ of the
pulsar is $5\,$Gyr.  Optical observations of the field have so far
failed to detect the companion, but set stringent limits, showing that
it must be an old, cold white dwarf (\cite{cch+89}; \cite{kdk91}).

Here, we report on Keck and {\it Hubble Space Telescope} observations
of the \psr\ field, in which the counterpart is detected.  We use
these to determine the effective temperature and discuss the
implications for cooling of helium white dwarfs and braking of
millisecond pulsars.

\section{Observations}\label{sec:obs}

The \psr\ field was observed on 10 August 1994 using the
Low-Resolution Imaging Spectrometer (\cite{occ+95}) on the 10\,m Keck
telescope.  Dithered exposures were taken with total integration times
of 40 minutes in~R and 33 minutes in~I.  The conditions were good,
with seeing of $\sim\!0\farcs8$ in~I and $\sim\!1\farcs1$ in~R.  These
images showed a faint object at the pulsar position, but because the
field is very crowded, it was not clear whether or not the object was
the result of a blend, and its magnitude was difficult to determine.

Therefore, the Wide Field Planetary Camera 2 aboard the {\it{}Hubble
Space Telescope} was used to observe the \psr\ field for one orbit
each on 6 August 1997 and 19 February 1998.  The first observation
consisted of three 700\,s exposures through the F555W filter (mean
wavelength $\bar\lambda=5397\,$\AA, effetive width
$\Delta\lambda=1226\,$\AA; \cite{hbc+95}), the second of five 260\,s
exposures through~F814W ($\bar\lambda=7924\,$\AA,
$\Delta\lambda=1500\,$\AA).  The field around the pulsar was put on a
clean spot on the CCD of the Planetary Camera~(PC).  Only the PC
images are used here.

\begin{figure*}
\centerline{\hbox{\psfig{figure=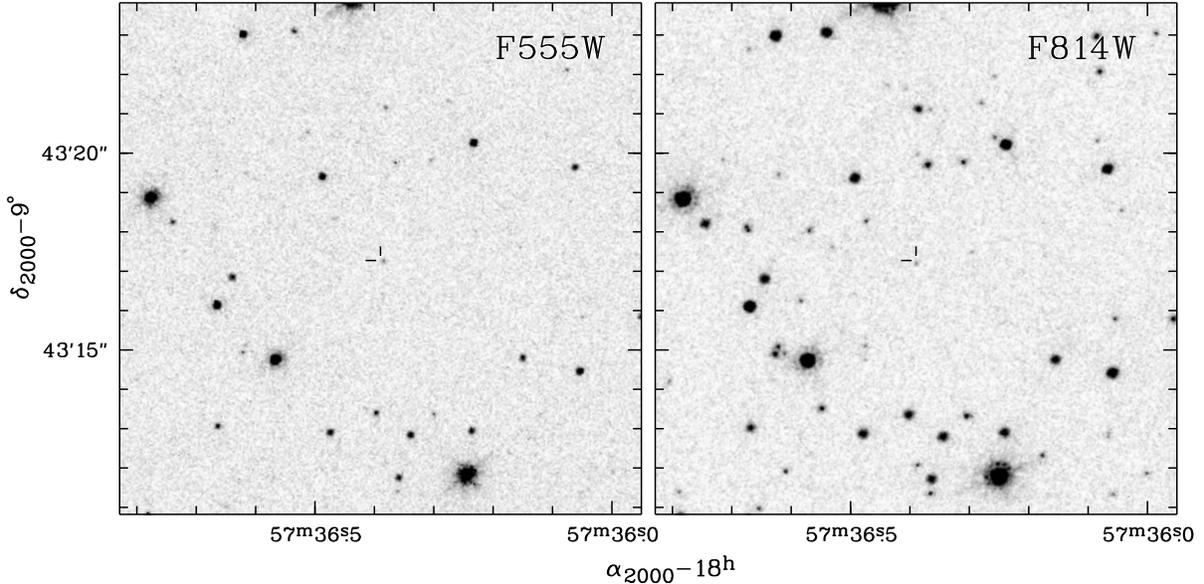,width=0.9\textwidth,angle=-90}}}
\caption[]{{\em HST\/} images of the \psr\ field, taken through the
F555W (left) and F814W (right) filters.  The epoch 1998.0 timing
position is shown by the tick marks.  These are 0\farcs24 long, equal
to the 95\% confidence diameter inferred from the uncertainty in the
astrometric tie.}
\label{fig:images}
\end{figure*}

For both data sets, the images were registered to half-pixel accuracy
and resampled using pixels with half the original size.  Next, cosmic
ray hits were identified by comparing values at a given position with
the minimum value occurring among the images at that position.  A
stacked image was formed using all unaffected pixels, and this image
was resampled to the original pixel scale for the further analysis.
The stacked images are presented in Figure~\ref{fig:images}.

Astrometry was done relative to the USNO-A2.0 catalog (\cite{mbc+98}),
using only those 189 objects that were not overexposed in a 10\,s
I-band Keck image and that appeared stellar and unblended.  We
measured their centroids and corrected for instrumental distortion
using a bi-cubic function determined by J.~Cohen (1995, private
communication).  Next, we fitted for zero-point position and position
angle on the sky, keeping the plate scale at the known value.
Rejecting 28 outliers (residuals $>\!1\arcsec$), the inferred
single-star measurement error is 0\farcs33 in each coordinate.  This
is somewhat larger than we have found for other fields (e.g.,
\cite{vkk99}), probably because in this crowded field there are
residual problems with blending in the USNO-A2.0 positions.

The solution was transferred to the F555W and F814W PC frames using 50
and 54 transfer stars, respectively.  For these, the PC positions were
corrected for instrumental distortion using the bi-cubic function
given by Holtzman et al.\ (\ctyr{hhc+95}).  We fitted for the
zero-point position only, as the plate scale and position angle on the
sky are known accurately (see \cite{hhc+95}).  Rejecting four and
three outliers for F555W and F814W, respectively (residuals
$>\!0\farcs08$), the inferred single-star measurement errors for both
are 0\farcs027 in each coordinate\footnote{Leaving scale and position
angle free, this reduces to 0\farcs014.  The final position, however,
changes by $<\!0\farcs006$.}.

The main uncertainty in our astrometric solution is the extent to
which the timing position of \psr\ and the USNO-A2.0 catalog are on
the same astrometric system.  The former is based on the DE200
dynamical ephemeris and should be close to the International Celestial
Reference System (ICRS; \cite{fcf+94}).  The USNO-A2.0 catalog is
tied to the ICRS as well (see \cite{mbc+98}).  There may be small
systematic differences, however, as well as effects of the average
proper motion of the USNO-A2.0 stars between the plate epoch
($\sim\!1954$) and the time of our Keck observation (1994.6) due to
Galactic rotation, asymmetric drift, and reflex of the solar motion.

We tried to measure the offset of our solution from the ICRS using two
Hipparcos (\cite{esa97}) stars, HIP~93083 and HIP~93084, which are on
our 10\,s I-band image.  These stars are strongly overexposed, but
nonetheless we were able to determine good centroids by fitting all
unaffected pixels, with different fitting methods agreeing to within
0\farcs03.  The offsets in both right ascension and declination
between the positions inferred from our solution and the two Hipparcos
stars (at epoch 1994.6) were consistent to within 0\farcs04
(0.02\,pix), the averages being $0\farcs17\pm0\farcs05$ and
$0\farcs34\pm0\farcs05$, respectively.

For verification, we took 20 AGN with VLBI-ICRS positions within
45\arcdeg\ of \psr\ from the list of Ma et al.\ (\ctyr{mef+98}) and
measured the positions of their optical counterparts on the second
Digitized Sky Survey (epoch $\sim\!1991$), with astrometry tied to
USNO-A2.0 (epoch $\sim\!1954$).  We find average offsets consistent
with the above, of $0\farcs18\pm0\farcs04$ and $0\farcs28\pm0\farcs04$
(the average offsets between the VLBI and the USNO-A2.0 positions are
0\farcs09 and 0\farcs11; the additional difference is consistent with
what is expected from the average proper motion of reference stars at
$\sim\!2\,$kpc).  We conclude that, after correction for the offset
found using the Hipparcos stars, our astrometry should be on the ICRS
to~0\farcs05.

Near the position of \psr\ -- $\rm\alpha_{J2000}=18^h57^m36\fs3917$,
$\rm\delta_{J2000}=09\arcdeg43\arcmin17\farcs275$ at epoch 1998.0
(\cite{ktr94}) -- the PC images show one object.  It is offset by
$-0\farcs01\pm0\farcs05$ and $-0\farcs04\pm0\farcs05$ in right
ascension and declination, respectively (see Fig.~\ref{fig:images}).
Given the density of objects of about 1 per three square arcseconds,
the probability of a chance coincidence in our 0\farcs12 radius 95\%
confidence error circle is~$\sim\!1.5\%$.  Thus, most likely we have
detected the counterpart of \psr.

Photometry of the object was done following the prescription of
Holtzman et al.\ (\ctyr{hbc+95}).  We performed aperture photometry
for a range of different radii, and used some two dozen brighter stars
in the frame to determine aperture corrections relative to the
standard 0\farcs5 (11\,pix) radius aperture.  For the candidate, the
best signal-to-noise ratio is for relatively small apertures, with
radii between 1.5 and 3\,pix.  From these, we infer a count rate for
the 0\farcs5 radius aperture of $0.042\pm0.005$ and
$0.087\pm0.009{\rm\,DN\,s^{-1}}$ for the F555W and F814W filters,
respectively ($1\,$DN is approximately 7 detected photons;
\cite{hbc+95}).  These count rates correspond to magnitudes of
$\mv=25.90\pm0.12$ and $\mi=24.19\pm0.11$ in the Vega system (using
$\mv=22.545$ and $\mi=21.639$ for a count rate of $1{\rm\,DN\,s^{-1}}$
in the PC, and applying a $0.10\,$mag aperture correction from
0\farcs5 radius to ``nominal infinity''; \cite{bcgr97}).

\section{Discussion}\label{sec:disc}

The temperature of the white dwarf can be constrained using the
measured color, $\mv-\mi=1.71\pm0.16$.  For this purpose, we correct
for reddening using the estimate $E_{B-V}=0.5\pm0.1$ of Kulkarni et
al.\ (\ctyr{kdk91}).  This estimate is based on CO emission and
\ion{H}{1} 21-cm absorption measurements towards \psr\ and is
consistent with the range 0.5--0.7 estimated in the general direction
of \psr\ from reddening of O--F stars (\cite{nks80}).  The redding
corresponds to $A_{\rm{}F555W}=1.6\pm0.3$ and
$E_{{\rm{}F555W}-{\rm{}F814W}}=0.65\pm0.13$ (\cite{sfd98}), and we
infer $(\mv-\mi)_0=1.06\pm0.21$.  This intrinsic color corresponds to
$\Teff=4800\pm800\,$K if the atmosphere were pure hydrogen
(\cite{bsw95}; using the color transformations of \cite{hbc+95}), but
could be lower than 4000\,K for a mixed helium/hydrogen atmosphere.
(Note that the surface gravity expected for the companion,
$\log{}g\simeq7.2$, is outside the range 7.5--8.5 covered by the
atmospheric models of Bergeron et al.\ [\ctyr{bsw95}], and thus we had
to extrapolate; the colors, however, are not very sensitive to
$\log{}g$.)

A consistency check on the estimated temperature is available using
the absolute magnitude.  From the measured white dwarf mass
($\Mcomp=0.258^{+0.028}_{-0.016}\Msun$; \cite{ktr94}), we infer a
radius $\Rcomp=0.021\Rsun$ (using the models of Driebe et al.\
[\ctyr{dsbh98}]; the result is not sensitive to model details).
Combined with our estimate of \Teff, one infers
$M_{\rm{}bol}\simeq13.9$, and, using bolometric corrections tabulated
by Bergeron et al.\ (\ctyr{bsw95}), $\Mv\simeq14.3$.  Correcting for
reddening, this corresponds to a parallax of 1.0\,mas, consistent with
the timing parallax of $1.1\pm0.3$\,mas.

With $\Teff$ and $\Mcomp$, the cooling age of the white dwarf can be
estimated using a cooling model.  As discussed in \Sref{intro}, the
main uncertainty in the models is the amount of nuclear burning, which
is set by the thickness of the hydrogen layer surrounding the helium
core.  Generally, hydrogen layers have been assumed to be relatively
thin, and little account has been taken of a possible dependence on
stellar mass.  For instance, Hansen \& Phinney (\ctyr{hp98}) modeled
helium white dwarfs having hydrogen layers of $3\times10^{-4}\Msun$
and $10^{-6}\Msun$.  With their models, one infers a cooling age for
the white dwarf here of $\sim\!3\,$Gyr.  Using simple evolutionary
models for the progenitor of the white dwarf, however, much thicker
layers are found; e.g., for a 1\,\Msun\ progenitor that ends up as a
0.259\,\Msun\ white dwarf, Driebe et al.\ (\ctyr{dsbh98}) find a
hydrogen layer of $4.8\times10^{-3}\Msun$ at formation.  With this
much thicker layer, nuclear burning is much more important, and the
cooling age becomes much longer, $\twd=10\pm2\,$Gyr (by which time
only $\sim\!0.7\times10^{-3}\,\Msun$ of hydrogen is left).

The white dwarf age inferred from the Driebe et al.\ model is greater
than the characteristic age of the pulsar, $\tauc=5\,$Gyr.  If the
model were correct, \tauc\ must be an underestimate of the true age of
the \psr\ system.  One interpretation is that the braking index of
this millisecond pulsar is less than the canonical value of~3; to
obtain $\tpsr>8\,$Gyr would require $n<2.25$ (Eq.~\ref{eq:tpsr}).
This is perhaps not unreasonable, as for most pulsars for which
braking indices have been measured, values less than~3 have been
found: $n=2.51\pm0.01$ for PSR~B0531+21 (\cite{lps93}); $2.28\pm0.02$
for PSR~B0540$-$69 (\cite{bvd+95}); $1.4\pm0.2$ for PSR~B0833$-$45
(\cite{lpsc96}); and $2.837\pm0.001$ for PSR~B1509$-$58
(\cite{kms+94}).  All these pulsars, however, are young and have
strong magnetic fields, so their relevance to the discussion here is
not clear.  We note that a variant on the vacuum-dipole model
(\cite{mel97}), which does a reasonable job of explaining these
braking indices, predicts $n=3$ for a pulsar like \psr\ (A.~Melatos \&
J.~Hibschman, 1999, private communication).

The optical counterparts of other pulsar binaries may give a clue to
where the problem lies.  From the list compiled by Hansen \& Phinney
(\ctyr{hp98}), we find that two pulsars, PSRs~J0034$-$0534 and
J1713+0747, have very cool companions, with $\Teff<3500\,$K and
$\Teff=3400\pm300\,$K, respectively.  For such temperatures, the
cooling ages inferred from the models of Driebe et al.\
(\ctyr{dsbh98}) are well over $10\,$Gyr, even if the orbital
inclinations were such that the helium white dwarfs had close to the
maximum mass\footnote{The companion of PSR~J0034$-$0534 could also be
a CO white dwarf, which would make the cooling age much shorter
(\cite{sdb99}).  The required low orbital inclination has $\sim\!10\%$
a priori likelihood.  The companion of PSR~J1713 has mass
$0.27<\Mcomp<0.4\,\Msun$ and thus should be a helium white dwarf
(\cite{cfw94}).}.  While these ages are not inconsistent with the
estimated pulsar ages, they exceed estimates of the age of the Galaxy
from the white-dwarf luminosity function (for recent determinations,
see \cite{lrb98}; \cite{khh99}).  This suggests that the models may
overestimate the cooling ages.

The above is in contrast to what is the case for PSR~J1012+5307, where
the cooling age estimated using the Driebe et al.\ model is very
similar to \tauc\ (\Sref{intro}).  The discrepancy might be resolved
by the thickness of the hydrogen layer being a function of the orbital
separation, perhaps via somewhat different mass-loss histories.
PSR~J1012+5307 has the second-shortest orbital period of all systems
known (0.6\,d), much shorter than that of \psr\ (12.3\,d).
PSR~J0034$-$0534, however, discussed above as another case for which
thick hydrogen layers may be problematic, has a short orbital period
(1.6\,d).

The discrepancy might also result from differences in white-dwarf
mass.  For instance, Driebe et al.\ (\ctyr{dbsh99}) find that shell
flashes only occur in a limited range of masses (0.21--0.30\,\Msun),
which includes \psr\ but not PSR~J1012+5307.  Driebe et al.\ found
that burning during the flashes does not greatly affect the cooling
ages.  Sch\"onberner, Driebe, \& Bl\"ocker (\ctyr{sdb99}), however, in
duscussing our result, have suggested that envelope expansion during
and following a flash could cause Roche-lobe overflow.  The resulting
mass transfer to the pulsar (which might temporarily become an X-ray
source again) could lead to a thinner hydrogen layer.  If so, then
with stronger constraints on companion masses, as can be derived for
PSR~J1713+0747 in particular, the mass range in which shell flashes
occur can be constrained observationally.

\acknowledgements We thank Brad Hansen, Thomas Driebe, and Thomas
Bl\"ocker for useful discussions, and the latter also for informing us
of new results prior to publication.  The observations were obtained
with the W. M. Keck Observatory on Mauna Kea, Hawaii, and the NASA/ESA
{\em Hubble Space Telescope}.  We acknowledge generous access to the
F814W images of \psr\ taken by Foster et al.\ (GO~6642), support of a
NASA Guest Observer grant, a fellowship of the Royal Netherlands
Academy of Arts and Sciences (MHvK), an Australian Research Council
Fellowship (JFB), an Alfred P. Sloan Fellowship (VMK), grants from
NASA and NSF (SRK), visitor grants of NWO and LKBF (VMK), and
hospitality of Utrecht University (VMK, JFB).

\end{document}